\documentstyle[pra,eqsecnum,aps]{revtex}

\def\displayfrac#1#2{\frac{\displaystyle #1}{\displaystyle #2}}

\begin{document}
\title
{Relativistic Operator Description of Photon Polarization}
\author{Arvind\\Department of Physics\\Indian Institute of Science, 
Bangalore 560012 \\and\\N.Mukunda\thanks{Also at Jawaharlal Nehru Centre for
Advanced Scientific Research, Jakkur, Bangalore 560064}\\Centre for
Theoretical Studies and Department of Physics\\ Indian Institute of
Science\\Bangalore 560012, India}
\maketitle
\begin{abstract}
We present an operator approach to the description of photon
polarization, based on Wigner's concept of elementary
relativistic systems.  The theory of unitary representations of
the Poincar\`{e} group, and of parity, are exploited to
construct spinlike operators acting on the polarization states
of a photon at each fixed energy momentum.  The nontrivial
topological features of these representations relevant for
massless particles, and the departures from the treatment of
massive finite spin representations, are highlighted and
addressed.
\end{abstract} 
\section{Introduction}
In the framework of relativistic quantum mechanics, elementary
systems are described by unitary irreducible representations
(UIR's) of the proper orthochronous inhomogeneous Lorentz group, or
Poincar\`{e} group $\cal{P}$\cite{1}.  This is the symmetry group of special
relativistic space-time.  The adjective `elementary' here implies that 
for such systems every physical observable can in principle be
constructed as a function of the operators implementing the
transformations of ${\cal P}$.  However, as is well known, not
all the mathematically constructible UIR's of ${\cal P}$ are
physically acceptable.  Only the finite mass, finite spin UIR's
containing positive time-like energy momenta, and the positive
lightlike energy momentum finite helicity UIR's, are realised in
nature.  These two classes of UIR's are closely linked to the
rotation subgroup $SO(3)$ of the homogeneous Lorentz group
$SO(3,1)$, and to an $E(2)$ subgroup of $SO(3,1)$, respectively.
UIR's of ${\cal P}$ containing space-like energy momenta, and the
light-like continuous spin or infinite helicity UIR's, are
unphysical\cite{2}.

On account of the special role played by the $SO(3)$ subgroup in
the timelike UIR's, there is a clean kinematic separation between
orbital and spin angular momenta in these representations.
Indeed the concept of spatial position is well-defined\cite{3}, and all
the generators of ${\cal P}$can be built up as functions of
kinematically independent position, momentum and spin operators.
Moreover in these UIR's of ${\cal P}$, action by the parity
operator $P$ can be accommodated without any enlargement of the
representation space.  Many of these simplifying features can
also be linked to the fact that the coset space $SO(3,1)/SO(3)$
(equivalently $SL(2,C)/SU(2)$) is ${\cal R}^3$ which is
topologically trivial.

In the lightlike finite helicity UIR's of ${\cal P}$, such as
one uses to describe photons, the situation is markedly
different in many respects.  Here the role of $SO(3)$ is taken
over by the $E(2)$ subgroup mentioned earlier, and this is not
directly linked to transformations on space alone.  Moreover the
coset space $SO(3,1)/E(2)$ is topologically nontrivial.  For
these reasons, there is no longer a clear cut kinematic
separation of anything analogous to spin from variables related
to space\cite{4}.  Indeed for nonzero helicity even position in space is
not a well-defined physical observable; and the parity operator
$P$ cannot be defined within such a UIR of ${\cal P}$.

On the other hand, beginning with classical optics, the
treatment of the polarization states of a plane electromagnetic
wave, and its extension to the possible polarization states of a
single photon, are both physically very well founded.  In polarization 
optics per se, intensity preserving linear optical systems - rotators, 
birefringent media, quarter wave plates, half-wave plates,...- are
effectively treated as realising various elements of an $SU(2)$ or
$U(2)$ group acting on the two-dimensional complex linear space
of polarization states\cite{5}.  However, from the point of view
of the theory of elementary systems, the relevant UIR's of
${\cal P}$ lead only to a single polarization state - helicity
$\pm~1$ corresponding to right or left circular polarization
respectively - and helicity is relativistically invariant.  One
must bring in the action by parity $P$ in an essential manner to
connect the states of opposite helicities and so create a
two-dimensional space of polarization states, allow for the
definitions of linear polarization, general elliptic
polarization etc.

The purpose of this paper is to examine this complex of
questions taking the relevant UIR's of ${\cal P}$ as a
fundamental starting point.  We wish to bring out the essential
role of parity in this context, something not often emphasized,
and develop the necessary operator machinery to deal with
transformations acting solely on the polarization degree of
freedom of a photon.  Since one deals here ultimately with an
irreducible representation of ${\cal P}$ extended by parity, in
principle all (physically important) operators can be built up
out of the generators of ${\cal P}$, finite transformations of
${\cal P}$ where necessary, and $P$.  We show that there is an
unavoidable momentum dependence in these constructions,
including in the building up of generators of $SU(2)$ acting on
the polarization states of a photon at fixed energy-momentum.

The contents of this paper are arranged as follows.  Section 2
sets up basic notational conventions for dealing with UIR's of
${\cal P}$, and includes the statement of the homomorphism from
$SL(2,C)$ to the homogeneous Lorentz group.  The use of
$SL(2,C)$ makes many later calculations much simpler than
otherwise.  The algebraic relations involving parity, the
Casimir invariants for ${\cal P}$, and the structure of the
positive energy time like UIR's with finite spin are briefly
reviewed for the convenience of the reader and for later comparison.
In particular the emergence of spin as a separately existing degree of
freedom in these UIR's, independent of space variables, is emphasized.  
In Section 3 we take up the mass zero finite helicity UIR's of 
${\cal P}$. Particular attention is paid to the nontrivial topological
features that emerge here, as compared to the finite mass case.
Alternative ways to pass from a standard energy-momentum to a
general energy momentum, in a singularity-free manner, and the
structure of Hilbert space basis vectors including their
transition rules, are developed.  Finally, the doubling of the
Hilbert space to accommodate parity, a feature absent in the
massive case, is described.  Section 4 shows how one can
construct an $SU(2)$ Lie algebra of operators, in a momentum
dependent way, to act on the polarization states of a photon for
each fixed energy momentum.  Here again the important role of
the parity operator $P$ is seen.  This construction too has to
be done avoiding singularities which would naively occur due to
the nontrivial topological features involved.  What emerges is
that there is no universal or global $SU(2)$ lying behind these
momentum dependent constructions, and at the same time there is
considerable freedom in the details of the constructions.
Section 5 contains some concluding remarks.

\section{Notations and the timelike UIR's of ${\cal P}$}
\setcounter{equation}{0}

We begin with some notational preliminaries.  The Lorentzian
metric will be chosen to be spacelike, with
$g_{\mu\nu}=\mbox{diag}(-1, 1, 1, 1),~\mu,\nu=0,1,2,3$.  The
two-to-one homomorphism from $SL(2,C)$ to $SO(3,1)$ is given as
follows:
     \begin{eqnarray}
     A\epsilon~SL(2,C):~~A\sigma_{\mu} A^{\dag}&=&\Lambda(A)^{\nu}~_{\mu}~
     \sigma_{\nu}~,\nonumber\\
     A~p\cdot \sigma A^{\dag} &=&(\Lambda(A)p)\cdot\sigma~,\nonumber\\
     p\cdot\sigma &=& p^0\cdot 1 +\underline{p}\cdot\underline{\sigma}~,
     \nonumber\\
     \Lambda(A)& \epsilon & SO(3,1)~,\nonumber\\
     \Lambda(A^{\prime})\Lambda(A) &=& \Lambda(A^{\prime} A)~\cdot
     \end{eqnarray}
\noindent
Here $\sigma_0=1$ and $\underline{\sigma}$ are the usual Pauli
matrices, and $p^{\mu}$ is any (real) four-vector.  In any
unitary representation (UR) or UIR of ${\cal P}$, we have ten
hermitian generators $M_{\mu\nu}=-M_{\nu\mu}, P_{\mu}$ obeying
the standard commutation relations
     \begin{eqnarray}
     \protect[M_{\mu\nu}, M_{\rho\sigma}\protect]&=&
     i(g_{\mu\rho}M_{\nu\sigma}-g_{\nu\rho}
     M_{\mu\sigma} +g_{\mu\sigma} M_{\rho\nu}-g_{\nu\sigma}M_{\rho\mu})~,
     \nonumber\\
     \protect[M_{\mu\nu}, P_{\rho}\protect]&=& 
     i(g_{\mu\rho} P_{\nu} - g_{\nu\rho} P_{\mu})~,  \nonumber\\
     \protect[P_{\mu}, P_{\nu}\protect]&=&0~\cdot
     \end{eqnarray}
\noindent
The six components of $M_{\mu\nu}$ generate homogeneous Lorentz
transformations, and the four $P_{\mu}$ generate space-time
translations.  In our work we will also have to deal directly
with unitary operators $\overline{U}(A)$ representing finite
elements of $SL(2,C)$.  Using the split notation
$M_{jk}=\epsilon_{jk\ell} J_{\ell}, M_{0j}=K_{j},
j,k,\ell=1,2,3$, we have for any real three-vectors
$\underline{\alpha},\underline{v}$ the identifications:
     \begin{eqnarray}
     \overline{U}\left(e^{i\underline{\alpha}\cdot\underline{\sigma}/2}\right)
     &=& e^{i\underline{\alpha}\cdot\underline{J}}~,\nonumber\\
     \overline{U}\left(e^{-\underline{v}\cdot\underline{\sigma}/2}\right)
     &=& e^{i\underline{v}\cdot\underline{K}}~,
     \end{eqnarray}
\noindent
and the general transformation law for $P^{\mu}$:
     \begin{eqnarray}
     \overline{U}(A)^{-1} P^{\mu} \overline{U}(A) =
     \Lambda(A)^{\mu}~_{\nu}~P^{\nu}~\cdot
     \end{eqnarray}
\noindent
Thus if $|p,\ldots>$ is an eigenstate of the energy momentum
operators $P^{\mu}$ with eigenvalues $p^{\mu}$, we have the
general rule (upto possible phases and normalisation)
     \begin{eqnarray}
     \overline{U}(A)|p,\ldots> &=& |p^{\prime},\ldots >~,\nonumber\\
     p^{\prime} &=& \Lambda(A) p~\cdot
     \end{eqnarray}

When the action by parity $P$ is defined, it has the following effects:
     \begin{eqnarray}
     P\overline{U}(A)P^{-1} &=& \overline{U}\left(A^{\dag -1}\right)~,
     \nonumber\\
     P\underline{J} P^{-1} &=& \underline{J}~,\nonumber\\
     P\underline{K} P^{-1}&=& -\underline{K}~,\nonumber\\
     P(P^0, \underline{P})P^{-1} &=& (P^0, -\underline{P})~\cdot
     \end{eqnarray}
\noindent
For a numerical four-vector $p=(p^0,\underline{p})$, we shall
always write $\tilde{p} =(p^0,-\underline{p})$.  Then to
accompany eqn. (2.5) we have, when $P$ is defined,
     \begin{eqnarray}
     P|p,\ldots\rangle = |\tilde{p},\ldots\rangle~\cdot
     \end{eqnarray}

Given any $UR$ of ${\cal P}$, the Pauli Lubanski pseudo vector
$W_{\mu}$ is defined by
     \begin{eqnarray}
     W_{\mu}&=&\frac{1}{2}\in_{\mu\nu\rho\sigma} M^{\nu\rho} P^{\sigma}~,
     \nonumber\\
     \in_{0123} &=& +1~,
     \end{eqnarray}
\noindent
and is orthogonal to $P^{\mu}$.  Then the two Casimir invariants
for ${\cal P}$ are
     \begin{eqnarray}
     {\cal C}_1 &=& -P^{\mu} P_{\mu}~,\nonumber\\
     {\cal C}_2 &=& W^{\mu} W_{\mu}~\cdot
     \end{eqnarray}
\noindent
In a UIR of ${\cal P}$, both reduce to numbers.

The positive energy timelike UIR's may be labelled by the pair $\{m,s\}$, 
where $m>0$ is the rest mass and $s=0,1/2,1,\ldots$ the intrinsic
spin.  The Casimir invariants have the values:
     \begin{eqnarray}
     \{m,s\} : {\cal C}_1 =m^2,~{\cal C}_2 = m^2 s(s+1)~\cdot
     \end{eqnarray}
\noindent
In this UIR, $P^{\mu}$ and $W^{\mu}$ are respectively positive
timelike and spacelike four-vectors with, however, noncommuting
components for $W^{\mu}$ (unless $s=0$, when $W_{\mu}$
vanishes).  The rest-frame energy-momentum, from which all others in 
this UIR can be
obtained by suitable Lorentz transformations, will be written as $p^{(0)}$:
     \begin{eqnarray}
     p^{(0)} =(m,0,0,0)~\cdot
     \end{eqnarray}
\noindent
The corresponding stability subgroup of $SL(2,C)$ is $SU(2)$:
     \begin{eqnarray}
     A p^{(0)}\cdot\sigma A^{\dag} = p^{(0)}\cdot \sigma \Leftrightarrow
     A = a~\epsilon~SU(2)~\cdot
     \end{eqnarray}
\noindent
An (ideal) basis of momentum eigenstates for the Hilbert space
${\cal H}(\{m,s\})$  carrying this UIR may be built up as
follows, starting with the vectors $|p^{(0)},s_3\rangle$
describing a spin $s$ particle at rest\cite{6}.
These vectors are characterised by their behaviour under
rest-frame rotations, ie., elements of $SU(2)$: 
     \begin{eqnarray}
     a~\epsilon~SU(2) :\overline{U}(a)\big|p^{(0)},s_3\rangle =
     \sum\limits_{s^{\prime}_{3}} D^{(s)}_{s^{\prime}_{3} s_{3}}(a)
     \big|p^{(0)}, s^{\prime}_{3}\rangle~,
     \end{eqnarray}

\noindent
where $D(a)$ is the unitary matrix representing
$a~\epsilon~SU(2)$ in the $(2s+1)$-dimensional spin $s~~~UIR$ of
$SU(2)$.  Going on now to general energy-momentum we have:
     \begin{eqnarray} p &=&(p^0,\underline{p}), p^0 =
     (m^2+\underline{p}^2)^{1/2}~:\nonumber\\ |p,s_3\rangle
     &=&\overline{U}(\ell(p))|p^{(0)},s_3\rangle, s_3=s, s-1,\ldots,
     -s~,\nonumber\\ \ell(p)&=&
     \displayfrac{1}{[2m(m+p^0)]^{1/2}}(m+p\cdot\sigma)\nonumber\\
     &=&\left(p+p^{(0)}\right)\cdot\sigma\big/
     [2m(m+p^0)]^{1/2}~,\nonumber\\
     \ell\left(p^{(0)}\right)&=&1~;\nonumber\\ \ell(p)
     p^{(0)}\cdot\sigma\ell(p)^{\dag} &=& p\cdot\sigma~;\nonumber\\
     P^{\mu}|p,s_3\rangle &=& p^{\mu}|p,s_3\rangle~;\nonumber\\
     \langle p^{\prime},s^{\prime}_3|p,s_3\rangle &=& p^0\cdot
     \delta_{s_{3}^{\prime}s_{3}}
     \delta^{(3)}\left(\underline{p}^{\prime}-
     \underline{p}\right)~\cdot 
     \end{eqnarray}

\noindent
A general vector $|\psi\rangle~\epsilon~{\cal H}(\{m,s\})$ has a
$(2s+1)$ -component momentum space wave function
$\psi_{s_{3}}(\underline{p})$ and squared norm given by
     \begin{eqnarray}
     \langle p,s_3|\psi\rangle &=& \psi_{s_{3}}(\underline{p})~,\nonumber\\
     \langle \psi|\psi\rangle &=&\int\displayfrac{d^{3}p}{p^{0}}
     \sum\limits_{s_{3}}|\psi_{s_{3}}(\underline{p})|^2~\cdot
     \end{eqnarray}
\noindent
The action of $\overline{U}(A)$ on $|p,s_3\rangle$ for general
$A\epsilon~SL(2,C)$ involves the Wigner rotation, an element of
$SU(2)$ acting on the spin projection $s_3$:
     \begin{eqnarray}
     \overline{U}(A)|p, s_3\rangle &=& \sum\limits_{s^{\prime}_{3}}
     D^{(s)}_{s^{\prime}_{3}s_{3}}(a(p,A))|p^{\prime},s^{\prime}_{3}\rangle~,
     \nonumber\\
     p^{\prime} &=& \Lambda(A) p~,\nonumber\\
     a(p,A) &=& \ell(p^{\prime})^{-1}~A~\ell(p)~\epsilon~SU(2)~\cdot
     \end{eqnarray}
\noindent
In case
$p=p^{(0)}$ and $A\epsilon~SU(2)$, we have the simplifications
$p^{\prime}=p^{(0)},~a\left(p^{(0)},~A\right)=A$, so we recover eqn.(2.13).

We notice (as is well known) that the pure Lorentz
transformation $\ell(p)$ given in eqn.(2.14) is globally
well-defined and singularity free for all
$\underline{p}\epsilon\;{\cal R}^3$.  Related to this is the
fact that in the UIR $\{m,s\}$ of ${\cal P}$, one can introduce
well-defined hermitian position, momentum and spin operators
$\underline{q},\underline{p}, \underline{S}$ out of which all
the generators $M_{\mu\nu},P_{\mu}$ of ${\cal P}$ can be
constructed.  The nonvanishing fundamental or primitive
commutation relations are:
     \begin{eqnarray}
     [q_j, p_k]=i\delta_{jk},~[S_j,S_k]=i~\epsilon_{jk\ell}S_{\ell}
     \end{eqnarray}
\noindent
In the momentum basis subject to the normalisation given in eqn.(2.14) we have 
     \begin{eqnarray}
     \underline{q}=i\left(\displayfrac{\partial}{\partial\underline{p}} -
     \displayfrac{1}{2}\displayfrac{\underline{p}}{(p^0)^2}\right)~\cdot
     \end{eqnarray}
\noindent
Starting from the irreducible set $\underline{q},\underline{p}, \underline{S}$
(where $\underline{S}$ generate the spin $s$ UIR of $SU(2))$ we
can reconstruct the generators of ${\cal P}$ via
     \begin{eqnarray}
     P^{\mu} &=&\left(\left(m^2+\underline{p}^2\right)^{1/2},~\underline{p}
     \right)~,\nonumber\\
     \underline{J} &=&\underline{q}~\land~\underline{p} +\underline{S}~,
     \nonumber\\
     \underline{K}&=& \frac{1}{2} \{p^0,\underline{q}\} +
     \displayfrac{\underline{p}\wedge\underline{S}}{m+p^0}~\cdot
     \end{eqnarray}
\noindent
Conversely, $\underline{q}, \underline{p}$ and $\underline{S}$
can be recovered from $M_{\mu\nu},P_{\mu}$.  This is the
so-called Shirokov-Foldy form for the generators of ${\cal P}$
in the UIR $\{m,s\}$\cite{7}.  (In the massless finite helicity UIR's,
however, no such clean separation of primitive dynamical
variables and generators is possible).  The action of parity,
$P$, can be taken to be
     \begin{eqnarray}
     P|p, s_3\rangle =\eta|\tilde{p}, s_3\rangle~,
     \end{eqnarray}
\noindent
where $\eta=\pm~1$ is the intrinsic parity.  No enlargement of
the representation space ${\cal H}(\{m,s\})$ is needed, and we
have consistency with eqns.(2.6.7).

\section{The Lightlike UIR's of ${\cal P}$ and the parity doubling}
\setcounter{equation}{0}

Now we turn to the mass zero finite helicity UIR's of ${\cal
P}$, to be denoted $\{0,\lambda\}$ with the helicity $\lambda=0,
\pm~1/2, \pm~1,\ldots$.  In such a UIR, $\lambda$ has a fixed
value; for definiteness we assume it is nonzero and integral.
For photons we need just $\lambda=\pm~1$.  In the UIR
$\{0,\lambda\}$ both Casimir operators ${\cal C}_1$ and ${\cal
C}_2$ vanish, while the pseudovector $W_{\mu}$ becomes a
multiple of $P_{\mu}$:
     \begin{eqnarray}
     \{0,\lambda\}~:\quad{\cal C}_1 &=& {\cal C}_2=0~,\nonumber\\
                         W_{\mu} &=&\lambda~P_{\mu}~\cdot
     \end{eqnarray}
\noindent
This explains why parity $P$ cannot be defined within the space
of a single UIR $\{0,\lambda\}$.  For the present we will work
with a single UIR with fixed $\lambda$, and at the end of this
Section turn to the question of accommodating parity.

Towards setting up a basis of energy-momentum eigenfunctions for
the Hilbert space ${\cal H}(\{0,\lambda\})$ carrying the UIR
$\{0,\lambda\}$, analogous to eqns.(2.14) for ${\cal
H}(\{m,s\})$, we begin by noting that the set $\sum$ of all
positive lightlike energy-momentum four-vectors,
     \begin{eqnarray}
     \Sigma =\left\{p~\epsilon~{\cal R}^4| p^{\mu} p_{\mu} = 0,~
     p^0>0\right\}
     \end{eqnarray}
\noindent
is topologically nontrivial, since it is essentially ${\cal
R}^3-\{\underline{0}\}\cite{8}$.  It is therefore convenient to express
$\Sigma$ as the union of two overlapping open subsets
$\Sigma_N,~\Sigma_S$, each of which is topologically trivial.
Using the light cone combinations $p_{\pm}=p^0\pm p_3$, we
define:
     \begin{eqnarray}
     \Sigma_N&=&\{p\epsilon\Sigma |p_+>0\}~,\nonumber\\
     \Sigma_S&=& \{p\epsilon\Sigma |p_->0\}~,\nonumber\\
     \Sigma &=&\Sigma_N\cup\Sigma_S~;\nonumber\\
     \Sigma_N\cap\Sigma_S&=&\{p\epsilon\Sigma |p_{\perp}=(p_1,p_2)
     \neq 0\}~\cdot
     \end{eqnarray}
\noindent
The subscripts $N,S$ indicate that the North pole on $S^2$ is included in
$\Sigma_N$, the South pole in $\Sigma_S$.

Now we need to choose a standard or fiducial energy-momentum
four-vector $p^{(0)}$, to replace the choice (2.11) in the time
like case. We take  $p^{(0)}$ to be
     \begin{eqnarray}
     p^{(0)} = (1,0,0,1)~\cdot
     \end{eqnarray}
\noindent
(No confusion is likely to arise in using the same symbol
$p^{(0)}$ as before).  Then we have:
     \begin{eqnarray}
     \tilde{p}^{(0)} &=& (1,0,0,-1)~;\nonumber\\
     p^{(0)}\epsilon~\Sigma_N,&\not{\epsilon}~\Sigma_S&;
     \tilde{p}^{(0)}~\epsilon~\Sigma_S,~\not{\epsilon}~\Sigma_N~\cdot
     \end{eqnarray}
\noindent
Indeed, the $p$'s  omitted from $\Sigma_N(\Sigma_S)$ are all
positive multiples of $\tilde{p}^{(0)}\left(p^{(0)}\right)$.
The stability subgroup of $p^{(0)}$ is an $E(2)$ subgroup in
$SL(2,C)$:
     \begin{eqnarray}
     A~p^{(0)}\cdot\sigma A^{\dag} &=& p^{(0)}\cdot\sigma \Leftrightarrow
     \nonumber\\
     A &=& h(\varphi,\alpha)\epsilon~E(2)\subset SL(2,C)~,\nonumber\\
     h(\varphi,\alpha)&=& \left(\begin{array}{cc}e^{i\varphi/2}&\alpha\\
     0&e^{-i\varphi/2}\end{array}\right)~,\nonumber\\
     0\leq \varphi \leq 4\pi~,&& \alpha~\epsilon~{\cal C}~\cdot
     \end{eqnarray}
\noindent
The topological nontriviality of $\Sigma$ is the same as that of
the coset space $SL(2,C)/E(2)$, since $\Sigma\simeq
SL(2,C)/E(2)$.

In the space of the UIR $\{0,\lambda\}$ the fiducial energy
momentum eigenvector $|p^{(0)},\lambda\rangle$ is characterised
by the fact that it provides a one-dimensional representation of
$E(2)$:
     \begin{eqnarray}
     \overline{U}(h(\varphi,\alpha))|p^{(0)},\lambda\rangle =
     e^{i\lambda\varphi}|p^{(0)},\lambda \rangle~\cdot
     \end{eqnarray}

\noindent
In terms of the infinitesimal generators $\underline{J},
~\underline{K}$ of rotations and pure Lorentz transformations,
this means
     \begin{eqnarray}
     J_3|p^{(0)},~\lambda\rangle &=&\lambda|p^{(0)},~\lambda\rangle~,
     \nonumber\\
     (J_1+K_2)|p^{(0)},\lambda\rangle &=&
     (J_2-K_1) |p^{(0)},\lambda\rangle = 0~\cdot
     \end{eqnarray}

\noindent
These eqns.(3.7,8) are the replacements for the earlier
eqn.(2.13) in the timelike case.
Now we need to find, for each $p~\epsilon~\Sigma$, an $SL(2,C)$
element whose associated Lorentz transformation will carry
$p^{(0)}$ to $p$:  this will enable us to set up other
energy-momentum eigenvectors $|p,\lambda\rangle$, and so build
up a basis for ${\cal H}(\{0,~\lambda\})$, similar to
eqn.(2.14).  However in contrast to the timelike case this
cannot be done in a globally smooth manner for all
$p\epsilon~\Sigma$\cite{8}.  This again is a consequence of the
nontrivial topology of $\Sigma\simeq SL(2,C)/E(2)$.  The problem
has to be handled separately over each of $\Sigma_N,\Sigma_S$.
To prepare for this, we employ the usual spherical polar angles
$\theta, \phi$ on $S^2$ and define the unit vector
$\underline{n}(\theta,\phi)$ and an element
$a(\theta,\phi)\epsilon~SU(2)$ as follows:
   \begin{eqnarray}
   0\leq\theta\leq\pi~,&&0\leq\phi\leq 2\pi~:\nonumber\\
   \underline{n}(\theta,\phi)&=&\underline{n}(2\pi-\theta, \pi+\phi) =
   -\underline{n}(\pi-\theta, \pi+\phi)\nonumber\\
   &=& (\sin\theta\cos\phi,~\sin\theta\sin\phi,~\cos\theta)~;\nonumber\\
   a(\theta,\phi)&=&a(-\theta,\pi+\phi)
   =\exp\left[\frac{i\theta}{2}(\sigma_1\sin\phi-\sigma_2\cos\phi)\right]~
   \epsilon~SU(2)~\cdot\nonumber\\
   \end{eqnarray}
\noindent
We express a general $p$ as $p^0(1,\underline{n}(\theta,\phi))$
and see that $\Sigma_N, \Sigma_S$ correspond to
$0\leq\theta<\pi,~0<\theta\leq\pi$ respectively.  Whereas
$\underline{n}(\theta,\phi)$ is well-defined all over
$S^2,~a(\theta,\phi)$ is undefined at $\theta=\pi$ (south pole).
For a general $\underline{n}\epsilon~S^2$ we have
     \begin{eqnarray}
     &&a(\theta,\phi)\underline{n}\cdot\underline{\sigma}a(\theta,\phi)^{\dag}
     =\underline{n}^{\prime}\cdot\underline{\sigma}~,\nonumber\\
     \underline{n}^{\prime}&=&(\mbox{right handed rotation by angle}~
     \theta\;\mbox{about} (-\sin\phi,\cos\phi,0))\underline{n}~,\nonumber\\
     \end{eqnarray}
\noindent
so in particular we get the useful relation
     \begin{eqnarray}
     a(\theta,\phi)\underline{n}(\theta^{\prime},\phi)\cdot\underline{\sigma}
     a(\theta,\phi)^{\dag} =\underline{n}(\theta^{\prime}+\theta,\phi)\cdot
     \underline{\sigma}~\cdot
     \end{eqnarray}

Now a possible solution to the problem of constructing Lorentz
transformations connecting $p^{(0)}$ to all $p\epsilon~\Sigma$
is given by using separate boost and rotation factors in a
step-by-step manner:
     \begin{mathletters}
     \begin{eqnarray}
     p\epsilon\Sigma_N&:&\ell(p) = a(\theta,\phi)\exp\left(\frac{1}{2}\ln
     p^0\cdot\sigma_3\right)~,\nonumber\\
     &&\ell(p) p^{(0)}\cdot\sigma\ell(p)^{\dag} =p\cdot\sigma~;\\
     p\epsilon\Sigma_S&:&\ell^{\prime}(p)=a(\theta-\pi,\phi)
     \exp\left(-\frac{1}{2}\ln p^0\cdot\sigma_3\right) ~i\sigma_2~,
     \nonumber\\
     &&\ell^{\prime}(p)p^{(0)}\cdot\sigma \ell^{\prime}(p)^{\dag} = 
     p\cdot\sigma
     \end{eqnarray}
     \end{mathletters}
\noindent
(Once again, the use of the symbol $\ell(p)$ here should not
cause any confusion with its use earlier in Section 2).  In the
structure of $\ell^{\prime}(p)$, the purpose of the first factor
$i~\sigma_2$ is to switch $p^{(0)}$ to $\tilde{p}^{(0)}$, and
then the rest follows easily.  As is to be expected, in the
overlap $\Sigma_N\cap\Sigma_S,~\ell(p)$ and $\ell^{\prime}(p)$
differ by an $E(2)$ element on the right:
     \begin{eqnarray}
     p\epsilon\Sigma_N\cap\Sigma_S~:~ \ell^{\prime}(p) =
     \ell(p) h(2(\pi-\phi),~0)~\cdot
     \end{eqnarray}
\noindent
We also have the particular values
     \begin{eqnarray}
     \ell\left(p^{(0)}\right) &=& 1~,\nonumber\\
     \ell^{\prime}\left(\tilde{p}^{(0)}\right)&=&i~\sigma_2~\cdot
     \end{eqnarray}
\noindent
With the aid of these definitions we can set up a basis of
energy-momentum eigenvectors for ${\cal H}(\{0,\lambda\})$:
     \begin{eqnarray}
     p\epsilon\Sigma_N&:&|p,\lambda\rangle =\overline{U}(\ell(p))|p^{(0)},
     \lambda\rangle = \overline{U}(a(\theta,\phi))
     e^{-iK_3\ln p^0}\cdot|p^{(0)},\lambda\rangle~;\nonumber\\
     p\epsilon\Sigma_S &:&|p,\lambda\rangle^{\prime} =
     \overline{U}(\ell^{\prime}(p))|p^{(0)},\lambda\rangle =
     \overline{U}(a(\theta-\pi,\phi))\cdot
     e^{iK_3\ln p^0}\cdot e^{i\pi J_2}|p^{(0)},\lambda\rangle~;\nonumber\\
     p\epsilon\Sigma_N\cap\Sigma_S&:& |p,\lambda\rangle^{\prime}
     =e^{-2i\lambda \phi}|p,\lambda\rangle~;\nonumber\\
     &&P^{\mu}(|p,\lambda\rangle~\mbox{or}~|p,\lambda\rangle^{\prime}) =
     p^{\mu}(|p,\lambda\rangle~\mbox{or}~|p,\lambda\rangle^{\prime})~\cdot
     \end{eqnarray}
\noindent
The overlap or transition rule results from eqns.(3.7,13).
These definitions may be supplemented by the inner products
     \begin{eqnarray}
     \langle p^{\prime},\lambda|p,\lambda\rangle &=&
     p^0\delta^{(3)}\left(\underline{p}^{\prime}-\underline{p}\right)~,
     \nonumber\\
     ^{\prime}\langle p^{\prime},\lambda|p,\lambda\rangle^{\prime}&=&
     p^0\delta^{(3)}\left(\underline{p}^{\prime}-\underline{p}\right)~\cdot
     \end{eqnarray}
\noindent
It is always implied that in $|p,\lambda\rangle(|p,\lambda\rangle^{\prime})$
the argument $p$ is restricted to $\Sigma_N(\Sigma_S)$.  A
general $|\psi\rangle\epsilon{\cal H}(\{0,\lambda\})$ has a
single component momentum space wave function and squared norm
given by
    \begin{eqnarray}
    \psi(\underline{p})&=& \langle p,\lambda|\psi\rangle~,\nonumber\\
    \langle\psi|\psi\rangle &=& \int \displayfrac{d^3p}{p^0} |
    \psi(\underline{p})|^2~\cdot
    \end{eqnarray}
\noindent 
Here since the part of $\Sigma$ omitted in $\Sigma_N$ is a set
of measure zero as far as the integral is concerned, we have
used only the basis kets $|p,\lambda\rangle$.  Moreover,
starting from eqn.(3.7) and following through the definitions
(3.15), one checks that
     \begin{eqnarray}
     J_3|p^{(0)},\lambda\rangle &=& \lambda|p^{(0)},\lambda\rangle~;\nonumber\\
     \underline{J}\cdot\underline{p}/p^0|p,\lambda\rangle 
     &=& \lambda|p,\lambda\rangle~,\nonumber\\
     \underline{J}\cdot \underline{p}/p^0 |p,\lambda\rangle^{\prime}
     &=& \lambda|p,\lambda\rangle^{\prime}~\cdot
     \end{eqnarray}

For a general $A\epsilon~SL(2,C)$, the action of
$\overline{U}(A)$ on $|p,\lambda\rangle$ or
$|p,\lambda\rangle^{\prime}$ can be computed in a manner similar
to the steps leading to eqn.(2.16).  If $p^{\prime}=\wedge(A)p$
is in the overlap $\Sigma_N\cap \Sigma_S$, the result can be
expressed as a ``Wigner phase'' times
$|p^{\prime},\lambda\rangle$ or equally well as another ``Wigner
phase'' times $|p^{\prime},\lambda\rangle^{\prime}$; on the other
hand, if $p$ is a multiple of either $p^{(0)}$ or
$\tilde{p}^{(0)}$, the result can be written in only one way.
Since we do not need these results explicitly, we omit the
details.

The choices for $\ell(p),\ell^{\prime}(p)$ in eqns.(3.12) were
based on simple step-by-step constructions to lead from
$p^{(0)}$ to $p$.  Alternative choices, $\tilde{\ell}(p)$ and
$\tilde{\ell}^{\prime}(p)$ say, more in the spirit of the
expressions in (2.14) in the timelike case, are also available\cite{9}:
     \begin{eqnarray}
     p\epsilon\Sigma_N&:&\tilde{\ell}(p)=\displayfrac{1}{\sqrt{2p_+}}
     (1-\sigma_3+\sigma\cdot p)=\displayfrac{1}{\sqrt{2p_+}}
     \left(\tilde{p}^{(0)}+p\right)\cdot\sigma~;\nonumber\\
     p\epsilon\Sigma_S&:&\tilde{\ell}^{\prime}(p)=
     \displayfrac{1}{\sqrt{2p_-}}(1+\sigma_3-\sigma\cdot p)\sigma_1
     =\displayfrac{1}{\sqrt{2p_-}}\left(p^{(0)}-p\right)\cdot
     \sigma \sigma_1~;\nonumber\\
     p\epsilon\Sigma_N\cap\Sigma_S&:& \tilde{\ell}^{\prime}(p)=
     \tilde{\ell}(p) h(2(\pi-\phi), 2(1-p_3)/|p_{\perp}|)~\cdot
     \end{eqnarray}
\noindent
Again, these choices are $E(2)$-related to the earlier ones:
      \begin{eqnarray}
      \tilde{\ell}(p)&=&\ell(p) h\left(0,(1+1/p^0)\cdot\tan \theta/2\cdot 
      e^{-i\phi}\right)~,\nonumber\\
      \tilde{\ell}^{\prime}(p)&=&\ell^{\prime}(p) h\left(0,(1-1/p^0)\cdot\cot
      \theta/2\cdot e^{i\phi}\right)    
      \end{eqnarray}
\noindent
Since $0\leq\theta<\pi$ in the first case and $0<\theta\leq\pi$
in the second, these expressions are well-defined in their
respective domains.  Here the angle arguments in
$h(\ldots,\ldots)$ turn out to vanish, hence we get
\underline{exactly the same} energy-momentum eigenvectors as
before:
     \begin{eqnarray}
     \overline{U}(\tilde{\ell}(p))|p^{(0)},\lambda\rangle &=&
     \overline{U}(\ell(p))|p^{(0)},\lambda\rangle = |p,\lambda\rangle~,
     \nonumber\\
     \overline{U}(\tilde{\ell}^{\prime}(p))|p^{(0)},\lambda\rangle
     &=& \overline{U}(\ell^{\prime}(p))|p^{(0)},\lambda\rangle
     = |p,\lambda\rangle^{\prime}~\cdot
     \end{eqnarray}
\noindent
Having exhibited these alternative choices, we now revert to the earlier ones.

So far the analysis has been limited to the single UIR
$\{0,\lambda\}$ of ${\cal P}$ for a fixed value of $\lambda$.  To
accommodate parity $P$ we have to adjoin the inequivalent UIR
$\{0,-\lambda\}$ and work in the doubled Hilbert space ${\cal
H}(\{0,\lambda\})\oplus {\cal H}(\{0,-\lambda\})$.  This entails
bringing in additional basis vectors
$|p^{(0)},-\lambda\rangle,~|p, - \lambda\rangle$ and
$|p,- \lambda\rangle^{\prime}$ for ${\cal H}(\{0,-\lambda\})$
exactly as in eqn.(3.15).  In this extended space we fix the
action of $P$ by assuming $P^2=1$ and setting:
     \begin{eqnarray}
     P|p^{(0)}, \pm\lambda\rangle = |\tilde{p}^{(0)},\mp\lambda
     \rangle^{\prime}~ \cdot
     \end{eqnarray}
\noindent
We need not include an intrinsic parity factor $\eta$ here as was
done in eqn.(2.20) since in any case $P$ switches vectors in the
two subspaces ${\cal H}(\{0,\pm\lambda\})$.  One can now follow
through the consequences of eqn.(3.22)  by exploiting the basic
relations (2.6) and the constructions (3.15) in each subspace to
obtain:
     \begin{eqnarray}
     P|p,\pm\lambda\rangle &=& |\tilde{p},\mp\lambda\rangle^{\prime}~,\nonumber\\
     P|p,\pm\lambda\rangle^{\prime}&=& |\tilde{p},\mp\lambda\rangle~\cdot
     \end{eqnarray}
\noindent
To these we adjoin the helicity statements in the extended space:
     \begin{eqnarray}
     \underline{J}\cdot\underline{p}/p^0(|p,\pm\lambda\rangle~\mbox{or}~
     |p,\pm\lambda\rangle^{\prime}) =
     \pm\lambda(|p,\pm\lambda\rangle~\mbox{or}~
     |p,\pm\lambda\rangle^{\prime})~\cdot
     \end{eqnarray}
\noindent
For photons, we take $\lambda=1$: then $|p,\pm 1\rangle,~|p,\pm 1
\rangle^{\prime}$ correspond respectively to right and left
circular polarizations.

\section{$SU(2)$ generators on the polarization space}
\setcounter{equation}{0}

For fixed $p\epsilon~\Sigma$, the two polarization states
$|p,\pm\lambda\rangle$ (or $|p,\pm\lambda\rangle^{\prime}$) form
a basis for a two-dimensional polarization space.  The existence
of a group of $SU(2)$ transformations acting on this space is
obvious.  Our aim is to see how to construct the generators of
this $SU(2)$ out of the generators of ${\cal P}$ and parity $P$.
From eqn.(3.24) the helicity operator is already diagonal in
this space and acts like the third Pauli matrix $\sigma_3$.  We
need to build up analogues to $\sigma_1$ and $\sigma_2$.

Now we notice that parity $P$ switches helicity $\pm\lambda$ to
$\mp\lambda$, but at the same time changes
$p=(p^0,\underline{p})$ to $\tilde{p}=(p^0,-\underline{p})$.  We
must therefore supplement action by $P$ with a spatial rotation
by amount $\pi$, about some axis perpendicular to
$\underline{p}$, which will bring $-\underline{p}$ back to
$\underline{p}$ but leave helicity unaltered.  Here we face the
same topological problem which has appeared earlier in another
guise - it is impossible to find  a $\underline{p}$-dependent
(unit) vector perpendicular to $\underline{p}$, for all
$\underline{p}\epsilon~S^2$, in a singularity - free manner.
However such nonsingular choices are available on
$\Sigma_N,\Sigma_S$ separately.  We define $\underline{e}(p),
\underline{e}^{\prime}(p)$ as follows:
     \begin{mathletters}
     \begin{eqnarray}
     p\epsilon~\Sigma_N&:& \underline{e}(p)=\underline{n}\left(
     \frac{\pi}{2},0\right) -2\sin\frac{\theta}{2}\cos\phi~
     \underline{n}\left(\frac{\theta}{2},\phi\right)~,\nonumber\\
     &&|\underline{e}(p)| = 1,~~\underline{p}\cdot\underline{e}(p)=0~;\\
     p\epsilon~\Sigma_S&:&\underline{e}^{\prime}(p) =
     \underline{n}(\pi/2, 0) -2\cos\frac{\theta}{2}\cos \phi~
     \underline{n}\left(\frac{1}{2}(\pi+\theta),\phi\right)~,
     \nonumber\\
     &&|\underline{e}^{\prime}(p)|=1,~~\underline{p}\cdot\underline{e}
     ^{\prime}(p) = 0~\cdot
     \end{eqnarray}
     \end{mathletters}
\noindent
These are by no means unique but suffice for our purposes; each
is also unambiguously defined in the corresponding domain.
Starting now with $p\epsilon~\Sigma_N$ we develop:
    \begin{eqnarray}
    &&e^{i\pi\underline{e}(p)\cdot\underline{J}} P|p,\pm\lambda\rangle =
    e^{i\pi\underline{e}(p)\cdot\underline{J}}|\tilde{p},\mp\lambda\rangle
    ^{\prime}\nonumber\\
    &=& e^{i\pi\underline{e}(p)\cdot\underline{J}}
    \overline{U}(a(-\theta,\pi+\phi))\cdot
    e^{iK_3\ln p^0}\cdot e^{i\pi J_2}|p^{(0)},\mp\lambda\rangle\nonumber\\
    &=& e^{i\pi\underline{e}(p)\cdot\underline{J}}
    \overline{U}(a(\theta,\phi))e^{i\pi J_2}\cdot
    e^{-iK_3\ln p^0}\cdot |p^{(0)},\mp\lambda\rangle\nonumber\\
    &=&\overline{U}(a(\theta,\phi))\cdot\overline{U}\left(
    a(\theta,\phi)^{-1}\cdot
    e^{\frac{i\pi}{2}\underline{e}(p)\cdot\underline{\sigma}} \cdot
    a(\theta,\phi)\cdot e^{\frac{i\pi}{2} \sigma_2}\right)
    \cdot e^{-iK_3\ln p^0}|p^{(0)},\mp\lambda\rangle\nonumber\\
    \end{eqnarray}
\noindent
The $SU(2)$ element appearing here can be simplified after some
algebra and use of eqn.(3.11):
     \begin{eqnarray}
     &&a(\theta,\phi)^{-1}\cdot e^{\frac{i\pi}{2}\underline{e}(p)\cdot
     \underline{\sigma}}\cdot a(\theta,\phi)\cdot e^{\frac{i\pi}{2}
     \sigma_2}\nonumber\\
     &=& a(\theta,\phi)^{-1} i\underline{e}(p)\cdot\underline{\sigma}
      a(\theta,\phi) i\sigma_2\nonumber\\
     &=& -i\sigma_3 = e^{-\frac{i\pi}{2}\sigma_3}~\cdot
     \end{eqnarray}
\noindent
Using this in eqn.(4.2) we get:
     \begin{eqnarray}
     e^{i\pi\underline{e}(p)\cdot\underline{J}}
     P|p,\pm\lambda\rangle&=&\overline{U}(a(\theta,\phi))\cdot
     e^{-i\pi J_3-iK_3\ln p^0}\cdot |p^{(0)},\mp\lambda\rangle\nonumber\\
     &=& e^{i\pi\lambda}|p,\mp\lambda\rangle~\cdot
     \end{eqnarray}
\noindent
By analogous calculations for $p\epsilon~\Sigma_S$ we again find:
     \begin{eqnarray}
     e^{i\pi\underline{e}^{\prime}(p)\cdot\underline{J}}
     P|p,\pm\lambda\rangle^{\prime} = e^{i\pi\lambda}|p,\mp\lambda\rangle
     ^{\prime}~\cdot
     \end{eqnarray}
\noindent
Thus these operator expressions act essentially like the first Pauli matrix
$\sigma_1$ in the polarization space.

For photons we set $\lambda=1$.  At each fixed $p$, we may then
make the following identifications:
    \begin{eqnarray}
    \underline{J}\cdot\underline{p}/p_0&\longrightarrow&\sigma_3~,\nonumber\\
    -e^{i\pi\underline{e}(p)\cdot\underline{J}}&P\longrightarrow &\sigma_1~,
    \nonumber\\
    i\displayfrac{\underline{J}\cdot\underline{p}}{p^0}~e^{i\pi\underline{e}
    (p)\cdot\underline{J}}&P\longrightarrow &\sigma_2
    \end{eqnarray}
\noindent
Here for definiteness we assumed $p\epsilon~\Sigma_N$.  The
meaning is that we are working in a basis of circular
polarization states, and in that basis the Hilbert space
operators standing on the left reduce in their actions to the
Pauli matrices on the right.  Apart  from  factors of
$\frac{1}{2}$, these then are the generators of the group of
$SU(2)$ transformations familiar in polarization optics.

In the correspondence (4.6) only $\underline{J}$ and $P$ are to
be treated as Hilbert space operators while $p^{\mu}$ are
$c$-numbers.  One may wonder whether the latter could be
replaced by the operators $P^{\mu}$, and whether one would then
somehow obtain an $SU(2)$ algebra not tied down to basis states
$|p,\pm \lambda\rangle$ for specific $p^{\mu}$.  This however
does not work out due to operator ordering problems.  One is
obliged to \underline{first} pick some numerical $p^{\mu}$, and
then use the correspondence (4.6) \underline{only} for action on
the states $|p,\pm\lambda\rangle,~|p,\pm\lambda\rangle^{\prime}$.

\section{Concluding remarks}

In this paper we have focussed on the operator aspects of the
description of photon polarization states, taking as primary
inputs the concerned UIR's of ${\cal P}$,  the operators
available in such representations, and the parity operation.
This is in the spirit of the definition of elementary systems in
relativistic quantum mechanics.  We have highlighted the many
novel features that arise in the treatment of massless particles
as compared to massive ones, which can all be traced back to the
nontrivial topology of the coset space $SL(2,C)/E(2)$, or of the
positive light cone with tip removed.  We have emphasized the
crucial role played by the parity operation in being able to
create a two-dimensional polarization space, and in the
construction of operators realising the $SU(2)$ Lie algebra on
this space, at each fixed energy-momentum.  In a general sense, we can
say that while for electrons the operator description of spin precedes
the description in terms of spin states, for photons it is usually the
description of various polarization states that is physically
immediate.  We have tried here to supplement this by an operator description 
in as straightforward a manner as possible.

Our handling of the topological features involved, and avoidance
of singularities in expressions, leaves considerable freedom in
the choices of coset representatives
$\ell(p),~\ell^{\prime}(p)$, fields of vectors
$\underline{e}(p),~\underline{e}^{\prime}(p)$ perpendicular to
$\underline{p}$, etc.  What must be clear is that there is an
essential momentum dependence in these constructions, which
cannot be eliminated.  For each $p\epsilon~\Sigma$, we do have
an $SU(2)$ group acting on the corresponding polarization space;
however these various $SU(2)$'s are not representatives of any
single natural globally defined $SU(2)$ at all.  In particular
there is no relation to the geometrical group of rigid rotations
in physical space, as there is in the definition of spin for
massive particles.

This helps us answer a question which is not as naive as one may
at first imagine.  Suppose we have two photons with distinct
energy momenta $p,p^{\prime}$ respectively.  Can one treat their
separate two-dimensional polarization state spaces as though
they were like spin half particle states, couple the two photon
polarizations to ``total spins'' 1 or 0, and handle them just as
one would handle the spins of two electrons?  The answer is that
this is not physically well founded, since the $SU(2)$ groups
involved are momentum dependent; there is little meaning to the
action of ``one and the same $SU(2)$ element'' on both photon
polarizations on account of the conventions and freedoms involved
in identifying the $SU(2)$ generators for each $p$.


\begin{thebibliography}{99}
\bibitem{9}  These have been developed in refs. (2) and (8), though
the notations are slightly different.
\end{thebibliography}
\end{document}